\documentstyle[12pt]{article}

\begin{document}
\rightline{BNL-60340}
\rightline{ILL-(TH)-94-8}
\bigskip\bigskip

\begin{center} {\Large \bf Associated Production of Higgs and
Weak Bosons, with $H\rightarrow b\bar b$, at Hadron Colliders} \\
\bigskip\bigskip\bigskip\bigskip
{\large\bf A.\ Stange and W.\ Marciano } \\ \medskip Physics Department \\
Brookhaven National Laboratory \\ Upton, NY\ \ 11973 \\ \bigskip\bigskip
\bigskip {\large\bf S.\ Willenbrock} \\ \medskip Department of
Physics \\
University of Illinois \\ 1110 West Green Street \\  Urbana, IL\ \ 61801 \\
\bigskip \end{center} \bigskip\bigskip\bigskip

\begin{abstract}
We consider the search for the Higgs boson at a high-luminosity Fermilab
Tevatron
($\sqrt s=$ 2 TeV), an upgraded Tevatron of energy $\sqrt s =$ 3.5 TeV, and the
CERN Large Hadron Collider (LHC, $\sqrt s=$ 14 TeV), via $WH/ZH$ production
followed by $H\rightarrow b\bar b$ and leptonic decay of the weak vector
bosons.
We show that each of these colliders can potentially observe the standard Higgs
boson in the intermediate-mass range, 80 GeV $<m_H<$ 120 GeV.
This mode complements the search for and the study of the intermediate-mass
Higgs boson via $H\rightarrow \gamma\gamma$ at the LHC. In addition, it
can potentially be used to observe the lightest Higgs scalar of the minimal
supersymmetric model, $h$, in a region of parameter space not accessible to
CERN LEP II or the LHC (using $h\to \gamma\gamma,ZZ^*$).
\end{abstract}

\addtolength{\baselineskip}{9pt}

\newpage

\section{Introduction}

\indent\indent We recently completed an analysis of the prospects for
discovering the standard
Higgs boson at the Fermilab Tevatron, a $\sqrt{s} = 2$ TeV $p\bar p$
collider, assuming 1 $fb^{-1}$ of integrated luminosity (which requires the
Main Injector upgrade) \cite{SMW}.  Our
motivation for undertaking that study was a desire to exploit the Tevatron
to its full potential.  We showed that the most promising mode is $WH$
production, followed by $H\rightarrow b \bar b$ and leptonic decay of
the weak bosons \cite{GNY,GKSG}
\footnote{For a full set of references, see Ref.~\cite{SMW}.}.
We concluded that it may be possible, though challenging,
to detect a Higgs boson in the mass region $m_H= 60-80$ GeV
using these modes (a region that will also be explored by CERN LEP II). The
high
luminosity which will be provided by the Main Injector (${\cal L} =
10^{32}/cm^2/s$) together with the ability to detect secondary vertices from
$b$
quarks with a silicon vertex detector (SVX) make the search for the Higgs
boson at the Tevatron potentially viable over this limited mass range.

LEP II can search for a Higgs boson of mass up to $80 - 85$ GeV
\cite{LEP2}.  Beyond that, the
CERN Large Hadron Collider (LHC), a 14 TeV high-luminosity $pp$ collider
(${\cal L}\approx 10^{34}$), is planned to cover the Higgs-boson mass
range from 80 GeV up to 800 GeV \cite{ATLAS,CMS}.
However, the ``intermediate mass'' region, 80 GeV $< m_H <$ 120 GeV, is
particularly difficult for hadron supercolliders.  At the LHC, the search
for the intermediate-mass Higgs boson relies on the rare decay mode
$H\rightarrow \gamma\gamma$, which requires maintaining excellent photon energy
and angular resolution while running at full luminosity \cite{ATLAS,CMS}.
Furthermore, this signal becomes inaccessible if the branching ratio
of the Higgs boson to two photons is sufficiently suppressed below the
standard-model
value.  This can occur, for example, if the coupling of the Higgs boson to
bottom
quarks is enhanced with respect to the standard-model value, thereby increasing
the total width of the Higgs boson, which is dominated by $H\rightarrow
b\bar b$ in the intermediate-mass region. Such a scenario can be realized
for the light Higgs scalar (or the heavy Higgs scalar) of the minimal
supersymmetric model. We therefore return to our study
of the search for the Higgs boson via $WH/ZH$, followed by $H\rightarrow
b\bar b$, to ask whether this mode could be used at future hadron colliders
to search for a Higgs boson above the mass range that will be covered by LEP
II.  Even if an intermediate-mass Higgs boson can be
detected at the LHC via $H\rightarrow \gamma\gamma$, it
would still be valuable to observe $WH$ production, and the decay
$H\rightarrow b\bar b$, to explore the coupling of the Higgs boson to
weak vector bosons and bottom quarks.

Along with the LHC, we consider two possible upgrades of the Tevatron
collider at Fermilab.  The first is to increase the luminosity of the
machine to $10^{33}/cm^2/s$, and perhaps more.  This could be achieved
by adding several new rings of magnets, some or all in the Main Injector
tunnel, to increase the intensity of the antiproton source \cite{Snow,GJ}.
The second is to increase the energy of the machine along with the
luminosity. Of the various energy upgrades which one could consider, an
``energy doubler''
significantly increases the physics potential of the machine, can be
built on a relatively short time scale, and may
be complementary to the LHC in some ways \cite{Breck}.  This machine would
produce a large sample of top quarks and
extend the search range for new physics, e.~g., supersymmetry,
$Z^{\prime}$ bosons, etc. The installation of a new ring of
high-field magnets to replace the existing 4.4 Tesla magnets is required.
The machine we consider is a 3.5 TeV (7.7 Tesla magnets) $p\bar p$
collider.\footnote{Another possibility
being discussed is a 4 TeV (8.8 Tesla magnets) $p\bar p$ collider.  The
results for that machine are similar to those for the 3.5 TeV machine,
as will be quantified in a later footnote.}

The total cross sections for the various Higgs-boson production processes
at the Tevatron\footnote{The name ``Tevatron'' will be reserved for the
$\sqrt s=$ 2 TeV $p\bar p$ collider.} and the LHC can be found in
Refs.~\cite{SMW} and \cite{AACHEN}, respectively.
We show in Fig.~1 the various cross sections at the 3.5 TeV $p\bar p$
collider.  The QCD corrections to $gg\rightarrow H$ \cite{QCD,GSZ,DK}
(a factor of 2.1 in the $\overline{MS}$ scheme with $\mu=m_H$)
and $q\bar q\rightarrow WH/ZH$ \cite{HW} (a factor of 1.2 in the
$\overline{MS}$
scheme
with $\mu=M_{VH}$) are included.  The $gg\rightarrow H$ cross section is
increased by a factor of about $2.5-4$ over the corresponding Tevatron cross
section for $m_H= 60-200$ GeV.  The
$WH/ZH$ cross section is increased by a factor of about 2 over the
corresponding
Tevatron cross section (see Ref.~\cite{SMW}).

In section 2 we reconsider our previous analysis of $WH$ and $ZH$
production, followed by $H\rightarrow b\bar b$ and leptonic decay
of the weak bosons, for the Tevatron, the $\sqrt s =$
3.5 TeV $p\bar p$ collider, and the LHC. We list the number of events
obtained with 10 $fb^{-1}$ of integrated luminosity; this corresponds to
one ``year'' ($10^7 s$) of running at an instantaneous luminosity of
$10^{33}/cm^2/s$.  We also consider larger integrated luminosities, with the
caveat that $b$-tagging with high efficiency and purity, which is essential to
the extraction of the signal, may prove to be difficult at higher instantaneous
luminosities.  With a bunch spacing of 20 $ns$, a luminosity of
$10^{33}/cm^2/s$
yields about 1.5 interactions per bunch crossing, which is acceptable for
tagging secondary vertices with the SVX.  However, with the same bunch
spacing, a luminosity of $10^{34}/cm^2/s$ yields about 15 interactions
per bunch crossing.  It is not known how successful $b$-tagging with the
SVX will be in an environment with many interactions per bunch crossing.

In section 3 we consider the search for the lightest Higgs scalar, $h$, of the
minimal supersymmetric model, which contains two Higgs doublets.  There is a
region of parameter space in which the light Higgs scalar (or the
heavy Higgs scalar) has enhanced coupling to the $b$ quark, suppressing the
branching ratio of $h\rightarrow \gamma\gamma$ such that it is unobservable
at the LHC.  We show that the mode $Wh$, with $h\rightarrow b\bar b$,
can potentially be used to observe this Higgs boson over some of the region
where $h\rightarrow \gamma\gamma$ is unobservable.  This mode is therefore
complementary to the two-photon search mode for the light supersymmetric Higgs
scalar at hadron colliders.  In section 4 we present our conclusions.

\section{$WH/ZH$, with $H\rightarrow b\bar b$}

\indent\indent Associated production of Higgs and weak vector bosons, with
$H\to b\bar b$, was considered in detail at the Tevatron in our previous
study \cite{SMW}.  The
analysis here closely follows that work, and we refer the reader to that
paper for additional details. The weak bosons are detected via their leptonic
decays ($W\to \ell\nu, Z\to \ell\ell$). The cuts made to simulate
the acceptance of the detector
are listed in Table 1. The jet energy resolution is taken to be
$\Delta E_j/E_j = 0.80/\sqrt {E_j} \oplus 0.05$, which
corresponds to a two-jet invariant-mass resolution of about
$\Delta M_{jj}/M_{jj} = 0.80/\sqrt {M_{jj}} \oplus 0.03$ (added
in quadrature).\footnote{The two-jet invariant-mass resolution may be degraded
somewhat due to semileptonic $b$ decays, which occur in $40\%$ of the events.}
We integrate the background over an invariant-mass bin of $\pm 2\Delta M_{jj}$;
for $M_{jj}=$ 100 GeV, this amounts to a bin width of 34 GeV.
A change from our previous analysis is that we reduce the $p_T$ threshold
for observing charged leptons to $p_{T}>10$ GeV (although we continue to
trigger on leptons of $p_{T}>$ 20 GeV).
This is important for rejecting the top-quark
background to the $WH$ signal. We also extend the
coverage for jets out to a rapidity of 4, for the same reason. We will
comment on the effect of reducing the jet coverage to a rapidity of 2.5
at the Fermilab machines.

\begin{table}[ht] \begin{center} \caption[fake]{Acceptance cuts used
to simulate the detector. The $p_{T\ell}$ threshold is greater for charged
leptons which are used as triggers (in parentheses).} \bigskip \begin{tabular}
{ll}
$|\eta_b|<2$ & $p_{Tb}>15$ GeV \\
$|\eta_\ell|<2.5$ & $p_{T\ell}>10$ GeV (20 GeV)\\
$|\eta_j|<4$ & $p_{Tj}>15$ GeV \\
$|\Delta R_{b\bar b}|>0.7 $ & $|\Delta R_{b\ell}|>0.7$ \\
${\not \!p}_{T}>20$ GeV & (for $W\to\ell\bar\nu$) \\
\end{tabular} \end{center} \end{table}

The cross sections for the signals and backgrounds at the Tevatron,
the 3.5 TeV $p\bar p$ collider, and the LHC, including all branching ratios
and acceptances, are shown in Figs.~2, 3, and 4, respectively.
The $WH/ZH$ cross sections at the 3.5 TeV collider and the LHC
are increased by factors of about 2 and 4, respectively, after acceptance cuts,
over the corresponding cross sections at the Tevatron.
The dominant irreducible backgrounds are
$Wb\bar b/Zb\bar b$ and, for $m_H$ near $M_Z$,  $WZ/ZZ$ with
$Z\rightarrow b\bar b$.
The dominant reducible background is $Wjj/Zjj$ (not shown in the
figures).\footnote{The $Wjj$ and $Zjj$ backgrounds were calculated using the
code
developed in Ref.~\cite{BHOZ}.}
In our previous
study, we demanded that at least one jet be identified as a $b$ jet to reduce
this background; it nevertheless remained the dominant background, assuming a
1\% misidentification of a jet as a $b$ jet.  When one considers higher
luminosity, one obtains enough $WH/ZH$ signal events that
it becomes advantageous to demand that {\it both} jets be identified
as $b$ jets.  This greatly reduces the $Wjj/Zjj$ background. At the
Fermilab colliders, this background is significantly less than the
irreducible  $Wb\bar b/Zb\bar b$ background, while at the LHC the $Wjj$
background is comparable to the $Wb\bar b$ background, and $Zjj$ is
negligible compared with $Zb\bar b$. The $Wjj/Zjj$ background is
relatively larger at the LHC because it is initiated mostly by gluon-quark
collisions, while the signal arises from quark-antiquark annihilation.
Similarly, the $Zb\bar b$ background at the LHC is initiated by gluon-gluon
collisions, which accounts for its relatively large size.
Assuming an efficiency $\epsilon_b=$ 0.3 for tagging a $b$ jet within the
fiducial volume of the SVX and with $p_T >$ 15 GeV, we use
$\epsilon_b^2=$ 0.09 as an estimate of the efficiency for a double $b$ tag.

Another reducible background to
the $Wb\bar b$ signal is from $t\bar t$ production, followed by
$t\bar t \rightarrow b\bar b W^+W^-$ with one $W$ missed. Reduction of this
background requires coverage of leptons to
small $p_{T}$ and jets to high rapidity, to reduce the likelihood of
missing a $W$ boson. We reject events with
a hadronic decay of the additional $W$ boson if a jet from the $W$ decay with
$p_T >$ 30 GeV, or two jets with $p_T>$ 15 GeV (and invariant mass near
$M_W$), are observed. We calculate that
the signal is accompanied by jets exceeding these cuts only about $10\%$
of the time at the Fermilab machines, and $30\%$ of the time at the
LHC \cite{HW}.\footnote{We do not reduce the signal to account for the
rejection of events due to jet radiation.  The backgrounds will also
have jet radiation, and it is beyond the scope of this work
to include these effects.}
We also impose the requirement that the transverse mass
of the trigger charged lepton plus the missing $p_T$ be less than $M_W$,
which is almost always true for the signal.
The resulting cross sections for
$m_t=$ 170 GeV (the approximate central value from precision electroweak
experiments \cite{TOP}) are shown in Figs.~2 -- 4.\footnote{The $t\bar t$
background in Fig.~2(a) is reduced from
our previous study, Fig.~3(a) of Ref.~\cite{SMW}, due to the increased coverage
for leptons and jets assumed of the detector.}   At the Fermilab colliders,
the $t\bar t$ background to the $WH$ signal never exceeds the $Wb\bar b$
background, and is significant only for $m_H>$ 100 GeV at the 3.5 TeV collider.
At the LHC, $t\bar t$ is the biggest background to the $WH$ signal for
$m_H>$ 80 GeV.
The $t\bar t$ background at the LHC is relatively large because it is
initiated by gluon-gluon collisions, while the
signal arises from quark-antiquark annihilation.
This background increases at the Fermilab colliders and the
LHC by factors of roughly 2 and 1.7, respectively, for every 20 GeV decrease
in the top-quark mass. At the Fermilab colliders, about
$60\%$ of the remaining $t\bar t$ background events are
from the leptonic decay of the missed $W$ boson, and $40\%$ from the hadronic
decay, for $m_H=$ 100 GeV; the ratio drops to 50/50 for $m_H=$ 140 GeV.
Decreasing the jet coverage to a rapidity of 2.5 increases the hadronic
contribution by about $50\%$, thus increasing the net $t\bar t$ background by
only about $20-25 \%$.
At the LHC, the ratio of leptonic to hadronic missed-$W$ events is about
70/30 at $m_H=100$ GeV, decreasing to about 67/33 at $m_H=$ 140 GeV.

There are other sources of top quarks at hadron colliders which can
contribute to the $WH/ZH$ background.  Single-top-quark production via
$W$-gluon fusion, $Wg\rightarrow t\bar b$, in which the ``initial'' (virtual)
$W$ boson is radiated from an incoming quark, yields a
$Wb\bar b + q$ final state after the top quark decays \cite{WG}.\footnote{We
thank Chris Hill for bringing the single-top-quark backgrounds to our
attention.}  We suppress this
background by rejecting events in which the jet formed from the
outgoing quark has rapidity less than 4 and $p_T>$ 30 GeV.
The resulting background is shown in Figs.~2 -- 4; it is comparable to the
$t\bar t$ background.  The $Wg\to t\bar b$ background increases at the Fermilab
colliders and the LHC by factors of roughly 1.4 and 1.2, respectively, for
every
20 GeV decrease in the top-quark mass.

Single-top-quark production can also occur via the weak process
$q\bar q\rightarrow t\bar b$, which again yields a $Wb\bar b$ final state
after the top quark decays.  This is an irreducible background, since there
are no additional particles in the final state.  This background is also
shown in Figs.~2 -- 4; it is comparable to the other top-quark backgrounds
at the Fermilab machines, but relatively much smaller at the LHC. It increases
at the Fermilab colliders and the LHC by factors of roughly 1.7 and 1.5,
respectively, for every 20 GeV decrease in the top-quark mass.

\begin{table}[hbt]
\caption[fake]{Number of signal and background events at the Tevatron, the
$\sqrt s =$ 3.5 TeV $p\bar p$ collider, and the LHC, per 10 $fb^{-1}$ of
integrated luminosity,
for production of the Higgs boson in association with a weak vector boson,
followed by $H\to b\bar b$ and $W\to \ell\bar \nu$,
$Z\to \ell\bar \ell$. The statistical significance of the
signal, $S/\sqrt B$, is listed in the last column for the $WH/ZH$ processes.
The cuts made to simulate the
acceptance of the detector are listed in Table 1.  We assume a $30\%$
efficiency
for detecting a secondary vertex per $b$ jet, within
the rapidity coverage of the vertex detector and with $p_{Tb}>15$ GeV.
We demand that both $b$ jets be identified. We also assume
a $1\%$ misidentification of light-quark and gluon jets as a $b$ jet. The
$t\bar t\rightarrow b\bar bW^+W^-$ background to the $WH$ signal is reduced
by rejecting events with an additional $W$ boson, which is identified either
via a charged lepton, a jet of $p_T>$ 30 GeV, or two jets with $p_T>$ 15 GeV.
The $Wg\rightarrow t\bar b$ background is reduced by rejecting events with an
additional jet of $p_T>$ 30 GeV.}
\bigskip
\begin{center}
\begin{tabular}{ccccccccc}
&&&Tevatron&2 TeV $p\bar p$&&&&\\
\\
$m_H$ (GeV) & $WH/ZH$ & $Wb\bar b/Zb\bar b$ & $WZ/ZZ$ & $Wjj/Zjj$ & $t\bar t$
& $Wg\rightarrow t\bar b$ & $q\bar q\rightarrow t\bar b$ & Signif. \\
\\
60  & 90/14   & 108/16  & -      & 17/2    & 3/-    & 4.7/- & 11/- & 7.5/3.4
\\
80  & 50/8.5  & 70/14   & 30/7.4 & 13/1.6  & 3/-    & 5.9/- & 16/- & 4.3/1.7
\\
90  & 37/6.5  & 59/12   & 45/11  & 11/1.4  & 4/-    & 5.8/- & 18/- & 3.1/1.3
\\
100 & 27/4.9  & 45/9.4  & 37/9   & 9/1.3   & 4/-    & 5.4/- & 20/- & 2.5/1.1
\\
120 & 14/2.5  & 31/6.7  & 3/0.72 & 6.5/0.9 & 5/-    & 4.4/- & 21/- & 1.7/0.87
\\
140 & 5.4/0.88& 20/4.5  & -      & 4.5/0.72& 5/-    & 3.2/- & 20/- &
0.74/0.38\\
\\
\\
&&&&3.5 TeV $p\bar p$&&&&\\
\\
60  & 138/23  & 183/48  & -      & 43/5.4  & 11/-   & 21/- & 19/- & 8.3/3.1  \\
80  & 81/14   & 121/41  & 59/13  & 33/5    & 16/-   & 26/- & 29/- & 4.8/1.8  \\
90  & 61/11   & 99/37   & 89/20  & 30/4.7  & 18/-   & 26/- & 33/- & 3.6/1.4  \\
100 & 47/8.6  & 82/30   & 73/17  & 26/4.1  & 19/-   & 25/- & 36/- & 2.9/1.2  \\
120 & 26/4.7  & 58/23   & 6/1.3  & 19/3.4  & 22/-   & 20/- & 40/- & 2.0/0.89 \\
140 & 10/1.8  & 43/17   & -      & 15/2.7  & 24/-   & 16/- & 40/- & 0.83/0.41\\
\\
\\
&&&LHC&14 TeV $pp$&&&&\\
\\
60  & 281/58  & 394/409 & -      & 186/44  & 291/- & 286/- & 42/-  & 8.5/2.7
\\
80  & 170/36  & 285/395 & 119/46 & 169/49  & 407/- & 392/- & 62/-  & 4.6/1.6
\\
90  & 132/28  & 237/369 & 178/68 & 159/47  & 448/- & 376/- & 70/-  & 3.4/1.3
\\
100 & 104/22  & 199/349 & 146/56 & 142/44  & 481/- & 364/- & 77/-  & 2.8/1
\\
120 & 60/13   & 145/276 & 12/4.5 & 115/38  & 529/- & 312/- & 84/-  & 1.7/0.68
\\
140 & 24/4.9  & 110/220 & -      & 94/31   & 552/- & 257/- & 83/-  &
0.72/0.31\\
\end{tabular}
\end{center}
\end{table}

We list in Table 2 the number of signal and background events per 10
$fb^{-1}$ of integrated luminosity, with a double $b$ tag,
at the Tevatron, the 3.5 TeV $p\bar p$ collider\footnote{At a 4 TeV $p\bar p$
collider, the $WH/ZH$ signal and
the backgrounds $Wb\bar b/Zb\bar b$ and $WZ/ZZ$ increase by about $15\%$
from the 3.5 TeV machine, increasing the significance of the signal by
roughly $7\%$.}, and the LHC, for a variety of Higgs-boson masses.
The statistical significance of the signal, $S/\sqrt B$, is listed in the
last column for the $WH/ZH$ processes.
If we define discovery of the $WH$ signal by the criterion of
a 5$\sigma$ significance, then with 30 $fb^{-1}$ of integrated luminosity
the reach of the Tevatron is about $m_H=$ 95 GeV, the reach of the 3.5 TeV
$p\bar p$ collider about 100 GeV, and the reach of the LHC also about 100 GeV.
The largest background at the Fermilab machines is $Wb\bar b$,
while at the LHC it is $t\bar t$ and $Wg\to t\bar b$. The $t\bar t$ cross
sections in Figs.~2 -- 4
represent a reduction of the $t\bar t$ cross section, before rejecting the
additional $W$ boson but with acceptance cuts, by a factor of about 1/35
at the Fermilab machines, and about 1/25 at the LHC.
If the rejection could be improved at the LHC, it would increase the
significance of the signal.  The corresponding rejection factor for the
extra jet in the $Wg\to t\bar b$ process is only a factor of about 1/3.  If the
top-quark mass proves to be closer to 190 GeV, the significance of the LHC
signal would increase by about $10\%$. If the misidentification probability
of a light-quark jet as a $b$ jet is increased from $1\%$ to $2\%$, the $Wjj$
background quadruples, but since it is not the largest background at any of the
machines, the statistical significance of the $WH$ signal drops by only about
$15\%$.

If integrated luminosities in excess of 30 $fb^{-1}$ can be achieved,
one can observe Higgs bosons of greater mass. With 100
$fb^{-1}$, the reach of the Tevatron is about 120 GeV, the 3.5 TeV $p\bar
p$ collider about 125 GeV, and the LHC about 120 GeV. Thus each machine
is potentially capable of covering the intermediate-mass region. However,
one must keep in mind the caveat in the Introduction regarding $b$ tagging
at high luminosity.

The significance of the $ZH$ signal is much less than that of the $WH$
signal due to the small branching ratio of the $Z$ boson to charged
leptons.\footnote{In our previous analysis at the Tevatron,
Ref.~\cite{SMW}, we included the decay of the $Z$ boson to neutrinos.
We do not include this mode in this analysis because of the ensuing reasons in
the text.}
If the decay of the $Z$ boson to neutrinos could be used as well,
it would increase the significance of the signal by a factor of roughly 2.4,
making it comparable to that of the $WH$ signal.  However, the $ZH$ signal
with $Z \to \nu\bar \nu$, $H \to b\bar b$, has no simple trigger, and
suffers from potentially large backgrounds from the QCD production of
$b\bar bj$ where the jet is either mismeasured or carries away significant
$p_T$
outside of the rapidity coverage of the detector.\footnote{We thank J.~Gunion
and T.~Han for this observation.}  It remains to be shown
whether a minimum missing $p_T$ threshold exists which reduces the
background while maintaining a significant fraction of the signal.
The observation of a signal in the $ZH$ channel would be valuable to confirm a
signal in the $WH$ channel.

The $WZ/ZZ$ background can be normalized by using the purely leptonic
decay mode.  However, the Higgs peak at $m_H=$ 90 GeV is about $75\%$ of
the $Z$ peak (for the $WH$ signal). Thus, if the Higgs-boson mass happens
to lie near the $Z$ mass, it will be
difficult to convince oneself that the observed peak in the $b\bar b$
invariant-mass spectrum is really the sum of the Higgs peak and the $Z$
peak, and not just the $Z$ peak with an underestimate of the $b$-tagging
efficiency.  If the Higgs mass is sufficiently different from the $Z$ mass,
one should observe both the Higgs and the $Z$ peaks. Better dijet
invariant-mass resolution would be helpful to separate the peaks, as well
as to increase the statistical significance of the signals.

If a Higgs boson is discovered via $WH$, with $H\to b\bar b$, the
determination of the Higgs-boson mass will be limited by the $b\bar b$
invariant-mass resolution.  However, a precise measurement of the
Higgs-boson mass can be made via its decay to two photons, if this mode
proves to be observable.

\section {The light Higgs scalar of the minimal supersymmetric model}

\indent\indent In a two-Higgs-doublet model, the couplings of
one or more of the physical Higgs bosons (two neutral scalars, $h$ and $H$;
a neutral pseudoscalar, $A$; and a charged scalar, $H^{\pm}$) to a given
fermion are generally enhanced \cite{HKS}.
The Higgs sector of the minimal supersymmetric model is a
two-Higgs-doublet model which is determined (at tree
level) by two parameters, conventionally chosen to be the ratio of the
Higgs-field vacuum-expectation values, $v_2/v_1 \equiv \tan \beta$, and the
mass of the neutral pseudoscalar, $m_A$ \cite{HHG}.  Most models employ
radiative
electroweak symmetry breaking \cite{IR}, which requires $\tan \beta> 1$,
ranging
up to $\tan \beta \approx m_t/m_b$, which one obtains in the simplest $SO(10)$
grand-unified
models \cite{ALS,ACK}. For $\tan \beta > 1$ the coupling of all the Higgs
bosons to bottom quarks is enhanced over the standard-model coupling.

There is an upper bound on the mass of the lightest Higgs scalar, $h$,
which depends on the top-quark
mass and the top-squark masses, as well as on other parameters which we
ignore for this discussion \cite{MASS}.  This upper bound is attained for
large $\tan \beta$, and is given by
\begin{equation}
m_h^2<M_Z^2 + \frac{3G_F}{\pi^2 \sqrt 2}m_t^4 \ln \frac{m_{\tilde t_1}
m_{\tilde t_2}}
{m_t^2}
\label{bound}
\end{equation}
where $m_{\tilde t_{1,2}}$ are the masses of the top squarks.  For
$m_{\tilde t_{1,2}} =$ 500 GeV,
the upper bound is $m_h<$ 107, 113, 121 GeV for $m_t=$ 150, 170,
190 GeV, respectively. For $m_{\tilde t_{1,2}} =$
1 TeV, the upper bound is $m_h<$ 115, 125, 138 GeV, respectively.

For a given value of $\tan \beta$, the upper limit on $m_h$ is approached
as $m_A\rightarrow \infty$, and the coupling of the $h$ to bottom quarks
(and to all other standard-model particles) becomes of standard-model strength.
This particle can be detected at the LHC via $h\rightarrow \gamma\gamma$ or
$h\rightarrow ZZ^{(*)} \rightarrow l^+l^-l^+l^-$.  However, for
$m_A \sim 100-200$ GeV and $\tan \beta >2$, the $h$ is too heavy to be
produced at LEP II, and its branching ratio to two photons is too suppressed
to be detected at the LHC. This leads to the well-known ``hole'' in the
$m_A, \tan\beta$ plane in which the $h$ (as well
as the other Higgs particles, $H$, $A$, and perhaps also $H^{\pm}$, depending
on the top-quark mass) cannot be detected by either LEP II or LHC (using the
above decay modes)
\cite{ATLAS,CMS,KZ,BBKT,GO,BCPS}.

The hole in the $m_A,\tan\beta$ plane is due to the enhanced
coupling of $h$ to bottom quarks, which suppresses the branching ratio of
$h\rightarrow \gamma\gamma$. It is therefore a natural place to make use of
the process $Wh$, with $h\rightarrow b\bar b$. The Tevatron, the
3.5 TeV $p\bar p$ collider, and the LHC
can potentially detect $Wh$,
with $h\rightarrow b\bar b$, as long as $h$ is not so heavy that its
branching ratio to $b\bar b$ is suppressed by its decay to $WW^{(*)}$.
For large $\tan\beta$, the coupling of $h$ to bottom quarks is sufficiently
enhanced that this branching ratio is close to its maximum value
of $92\%$ all the way up to the real $WW$ threshold.  Because of this,
larger Higgs-boson masses can be accessed in the minimal supersymmetric
model than in the standard model,
where the branching ratio to $b\bar b$ begins to fall off at $m_H\approx$
110 GeV (see Fig.~2 in Ref.~\cite{SMW}).

We show in Figs.~5, 6, and 7 the discovery reach for the lightest
supersymmetric
Higgs scalar via $Wh$, with $h\rightarrow b\bar b$, at the Tevatron, the
3.5 TeV $p\bar p$ collider, and the LHC, respectively, in the
$m_A, \tan\beta$ plane, assuming a.) $m_{\tilde t_{1,2}}=$ 500 GeV and
b.) $m_{\tilde t_{1,2}}=$ 1 TeV. The
contours indicate the number of $fb^{-1}$ of integrated luminosity needed to
yield a 5$\sigma$ signal (for $m_h>$ 50 GeV, the approximate lower
bound from LEP I \cite{ALEPH}).
To guide the eye, the regions corresponding to 50 $fb^{-1}$ or less have been
shaded.
The upper right-hand corner of each plot corresponds to $h$ at the
maximum value of its mass and with standard-model couplings, so it is as
difficult to detect as the standard Higgs boson of the same mass.
As the top-squark masses decrease, the upper bound on $m_h$
decreases, and more of the $m_A, \tan\beta$ plane is covered for a given
amount of integrated luminosity. At all three machines, $h$ can be
detected with 30 $fb^{-1}$ only in the region $\tan\beta < 4$.
With 50 $fb^{-1}$ of integrated luminosity, the hole where the
$h$ is not accessible to LEP II and the LHC (via $h\rightarrow \gamma\gamma,
ZZ^*$) is almost entirely filled at the 3.5 TeV machine for
$m_{\tilde t_{1,2}}=$ 500 GeV (Fig.~6(a)) and is partially filled at the
3.5 TeV machine for $m_{\tilde t_{1,2}}=$ 1 TeV (Fig.~6(b)) and the LHC for
$m_{\tilde t_{1,2}}=$ 500 GeV
(Fig.~7(a)).  With 100 $fb^{-1}$, almost all of the hole is filled at
all three machines. This demonstrates that the coverage of the "hole" region
is sensitive to small changes in the analysis.  The upper bound on $m_h$
increases as the top-quark mass
increases, but this is largely compensated by the decrease in the top-quark
backgrounds, such that the coverage of the $m_A, \tan\beta$ plane is
not very sensitive to the top-quark mass.
If decays of $h$ to pairs of supersymmetric particles are kinematically
available, they may decrease the
branching ratio of $h\rightarrow b\bar b$, although for large $\tan\beta$
this decay may still dominate.

It may also be possible to observe the heavy Higgs scalar, $H$, of the minimal
supersymmetric model via $WH$, with $H\rightarrow b\bar b$, if $H$ is
relatively light.  For a fixed value of $\tan\beta$, the heavy Higgs scalar
approaches its lower bound, which is the same as the upper bound on the
mass of the light Higgs scalar, Eq.~\ref{bound}, as $m_A \rightarrow 0$.
For large $\tan\beta$, there is a region of parameter space in which $H$ is
near its lower bound, $h$ is heavier than 50 GeV, the coupling of $H$ to weak
vector bosons is almost of standard-model strength, and its coupling to bottom
quarks is at least of
standard-model strength.  This region lies roughly between $m_A=$ 50 and
100 GeV for large $\tan\beta$, depending on the top-quark mass
and the top-squark masses.

\section{Conclusions}

\indent\indent The intermediate-mass Higgs boson, 80 GeV $ < m_H <$ 120 GeV, is
elusive. It is too heavy to be produced at LEP II, and can be discovered at the
LHC using the rare two-photon decay mode only if excellent photon energy and
angular resolution can be maintained while running at full luminosity
(${\cal L}\approx 10^{34}/cm^2/s$).  We have shown that Higgs-boson production
in
association with a weak vector boson, followed by $H\rightarrow b\bar b$
and leptonic decay of the weak vector boson, can potentially be used to
observe the standard Higgs boson in the intermediate-mass range at future
hadron colliders. The
machines we considered are the Fermilab Tevatron ($\sqrt s=$ 2 TeV) with high
luminosity (${\cal L}\geq 10^{33}/cm^2/s$), an
upgraded Tevatron of energy $\sqrt s=$ 3.5 TeV with high luminosity, and the
LHC
($\sqrt s=$ 14 TeV).  With 30
$fb^{-1}$ of integrated luminosity, a 5$\sigma$ signal is possible in the
$WH$ channel for a Higgs boson of mass up to about 95 GeV at the Tevatron,
about 100 GeV at the 3.5 TeV $p\bar p$ collider, and also about 100 GeV at
the LHC. With 100 $fb^{-1}$ of integrated luminosity, the reach of the
Tevatron,
the 3.5 TeV $p\bar p$ collider, and the LHC is about 120, 125, and 120 GeV,
respectively.  However, to gather this amount of integrated luminosity would
require $b$-tagging in an environment with many interactions per bunch
crossing, which may prove to be difficult.  We conclude that each of these
machines can potentially cover the intermediate-mass region.

If the coupling of the Higgs boson to bottom quarks is enhanced, the decay
of the intermediate-mass Higgs boson to two photons can be suppressed such
that it is unobservable at the LHC.  This can occur for the light
Higgs scalar, $h$, of the minimal supersymmetric model, whose mass lies
within or below the intermediate-mass region if the top quark and the top
squarks are not too heavy. The enhanced coupling of $h$ to bottom
quarks increases the mass at which the
branching ratio of $h\rightarrow b\bar b$ begins to fall off due to the
competition from $h\rightarrow WW^{(*)}$.  We showed that the mode $Wh$,
with $h\rightarrow b\bar b$, can potentially cover part of the parameter
space of the minimal supersymmetric model not accessible to LEP II or the LHC
(using $h\to \gamma\gamma,ZZ^*$).  It is also possible that the heavy Higgs
scalar of the minimal supersymmetric model, $H$, can be observed via $WH$
production over part of the parameter space.

We hope that this paper will revitalize interest in associated production
of Higgs and weak vector bosons, with $H\rightarrow b\bar b$, at future
hadron colliders.  We encourage the detector collaborations at the Tevatron
and the LHC to undertake a deeper study of the potential usefulness of
this mode.

\section{Acknowledgements}

We are grateful for conversations with M.~Albrow, D.~Amidei, J.~Butler,
S.~Dawson, K.~Einsweiler, D.~Errede, S.~Errede, D.~Finley, B.~Foster,
H.~Gordon, J.~Gunion, T.~Han, C.~Hill, I.~Hinchliffe, L.~Holloway, R.~Kauffman,
T.~LeCompte, T.~Liss, F.~Paige, S.~Protopopescu, X.~Tata, and H.~Weerts.
This work was supported under contract no. DE-AC02-76CH00016 with the
U.S. Department of Energy.

 \clearpage

\section{Figure Captions}

\vrule height0pt \vspace{-22pt}

\bigskip

\indent Fig.~1 - Cross sections for various Higgs-boson production processes at
a $\sqrt s = 3.5$ TeV $p\bar p$ collider, versus the Higgs-boson mass.
The HMRSB parton
distribution functions \cite{HMRS} are used for all calculations.  The
top-quark mass is taken to be 170 GeV.
\bigskip

Fig.~2 - Cross sections and backgrounds at the Tevatron ($\sqrt s=$ 2 TeV)
for a.) $WH$ and b.) $ZH$ production,
followed by $H\to b\bar b$ and $W\to \ell\bar \nu$, $Z\to \ell\bar
\ell$, versus the Higgs-boson mass. The cuts made to
simulate the acceptance of the detector are listed in Table 1. The backgrounds
are from $Wb\bar b$ and $Zb\bar b$; $WZ$ and $ZZ$ followed by $Z\to b\bar b$;
$t\bar t \to W^+W^-b\bar b$ with one $W$ missed; $Wg\rightarrow
t\bar b \to Wb\bar b$; and $q\bar q\to t\bar b\to Wb\bar b$ (the top-quark
backgrounds for the $WH$ signal only).
The top-quark mass is taken to be 170 GeV.

Fig.~3 - Same as Fig.~2, but for the $\sqrt s=$ 3.5 TeV $p\bar p$ collider.

Fig.~4 - Same as Fig.~2, but for the LHC ($\sqrt s=$ 14 TeV $pp$).

Fig.~5 - Contour plots in the $m_A, \tan\beta$ plane of the number of
$fb^{-1}$ of integrated luminosity
required to observe a 5$\sigma$ signal for the light Higgs scalar, $h$, of the
minimal supersymmetric model via the process $Wh$, with $h\rightarrow
b\bar b$, at the Tevatron ($\sqrt s=$ 2 TeV).  For a given value of the
pseudoscalar Higgs mass, $m_A$, and the ratio of the Higgs-field
vacuum-expectation values, $v_2/v_1\equiv\tan\beta$, the mass of $h$
depends on the top-quark mass and the top-squark masses.  The figures are for
a.) $m_{\tilde t_{1,2}}=$ 500 GeV and b.) $m_{\tilde t_{1,2}}=$ 1 TeV.
The top-quark mass is taken to be 170 GeV; however, the
figures are insensitive to the top-quark mass, as explained in the text.
The regions of 50 $fb^{-1}$ or less are shaded to guide the eye.

Fig.~6 - Same as Fig. 5, but for the $\sqrt s=$ 3.5 TeV $p\bar p$ collider.

Fig.~7 - Same as Fig. 5, but for the LHC ($\sqrt s=$ 14 TeV).

\bigskip

\vfill

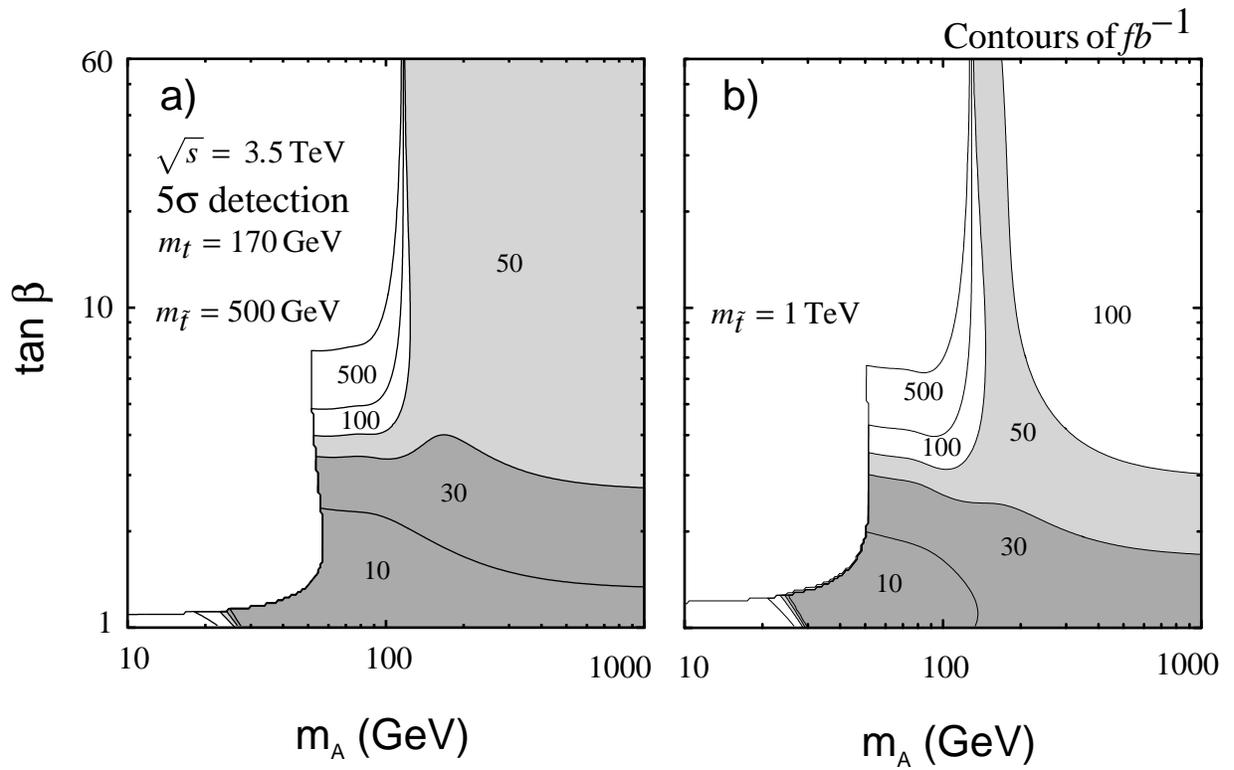
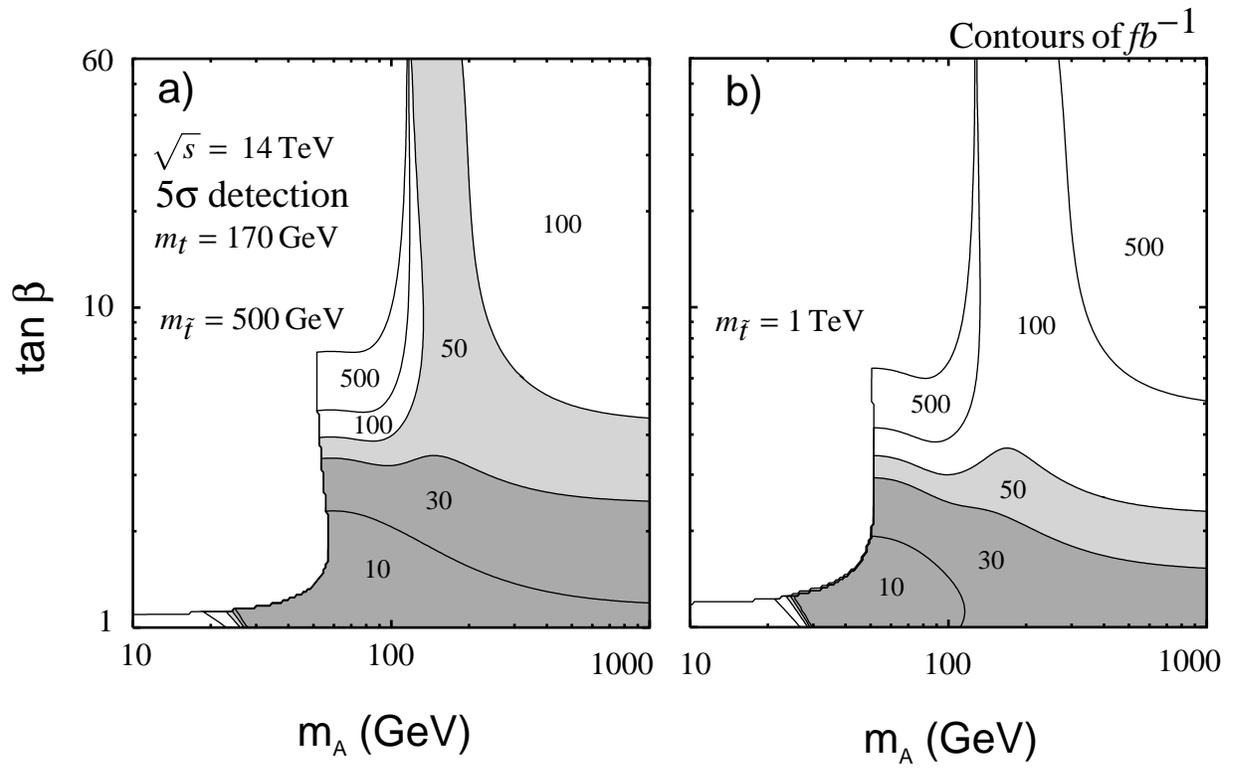
\end{document}